# Neutron-irradiation effects in LaO$_{0.9}$F$_{0.1}$FeAs superconductor

A. E. Karkin[1], J. Werner[2], G. Behr[2] and B. N. Goshchitskii[1]

[1]Institute of Metal Physics, Ural Branch of the Russian Academy of Sciences, 620219, Ekaterinburg, Russia
[2]Institute for Solid State and Materials Research IFW Dresden, 01069 Dresden, Germany

The effect of atomic disorder induced by neutrons irradiation on superconducting and normal state properties of polycrystalline LaFeAsO$_{0.9}$F$_{0.1}$ was investigated. The irradiation of the sample by a "moderate" neutron fluence $\Phi = 1.6 \cdot 10^{19}$ cm$^{-2}$ at $T_{irr} = 50 \pm 10$ °C leads to the suppression of superconductivity which recovers almost completely after annealing at temperatures $T_{ann} \leq 750$ °C. It is shown that the reduction of superconducting transition temperature $T_c$ under atomic disordering is not determined solely by the value of Hall concentration $n_H$, i.e. doping level, but is governed by the reduction of electronic relaxation time $\tau$. This behavior can be described qualitatively by universal Abrikosov-Gor'kov equation which presents evidence on the anomalous type of electrons pairing in Fe-based superconductors.

**Introduction**

The discovery of high-temperature superconductivity (SC) in the layered iron-based compounds [1] resulted in active experimental and theoretical studies of these systems. Full characterization of disordering effects can present new information on tentatively specific mechanism of electrons coupling in these superconductors [2]. The study of effect of atomic disordering induced by fast neutrons irradiation on electronic, magnetic and other subsystems allows obtaining new information on the properties of initially ordered systems. Of the special importance is the fact that the neutrons present an unique instrument to create the atomic-scale defects in solid without deviations from stoichiometry. The high penetrability of neutrons results in uniform distribution of the defects in bulk, which enables irradiating of macroscopic samples to be used in any other experiment.

The specific features of the radiation-induced disordering are its variability and reversibility. The former allows defects concentration to be varied in a wide range. The latter opens the possibility of gradual recovery of the sample initial state after high-temperature annealing. The high temperature treatment is accompanied by the successive recombination of different type defects. Thus, the same sample can be used for the series of measurements of physical properties with variation of neutron fluence and annealing temperature.

This method already shed new light on the physics of various superconducting systems. Fundamentally different with respect to each other superconducting properties were found in MgB$_2$ and YBa$_2$Cu$_3$O$_{7-\delta}$ under irradiation [3]. The magnesium diboride demonstrated relatively weak dependence of the critical temperature $T_c$ on neutrons fluence typical for the systems with strong electron-phonon interaction. The electron coupling here was found to be of an isotropic $s$-type. The yttrium-barium cuprate showed rapid decrease of critical temperature $T_c$ with neutron fluence growth. The fast and complete degradation of superconductivity in YBa$_2$Cu$_3$O$_{7-\delta}$ under irradiation presents evidence on different electron coupling mechanism.

Deep understanding of superconductivity mechanism in new class of FeAs-superconductors will require using various methods including study of radiation-induced disordering effects on kinetic and thermodynamic properties of these materials in normal and superconducting states. The available information on the transition temperature $T_c$ vs. residual resistivity $\rho_0$ dependence does not correlate with doping level of the samples [1, 4, 5, 6]. In part, this is due to the fact that the polycrystalline samples used in the majority of experiments were characterized by different contents of impurity phases, grain boundaries and doping elements.

In present work, the effects of atomic disordering induced by fast neutron irradiation of fluence $\Phi = 1.6 \cdot 10^{19}$ cm$^{-2}$ on kinetic properties of LaFeAsO$_{0.9}$F$_{0.1}$ in normal and superconducting states are investigated. The processes of recovery of initial properties of these materials as a result of multiple annealing of fixed duration at various temperatures are studied.

**Experimental**

Polycrystalline sample of LaO$_{0.9}$F$_{0.1}$FeAs consisting of 1–100 μ sized grains was prepared by using a two-step solid state reaction and annealed in vacuum [7]. The resistivity $\rho$ and the Hall coefficient $R_H$ were measured on the plate-like sample with dimensions $2.0 \times 1.5 \times 0.4$ mm$^3$ using the standard four-point method [8] with changing directions of probing current and magnetic field and switching between the current and potential leads. Measurements were done in the temperature range $T = 1.5 – 380$ K in magnetic field up to 13.6 T. The sample was irradiated by neutrons fluence $\Phi = 1.6 \cdot 10^{19}$ cm$^{-2}$ at $T_{irr} = 50 \pm 10$ °C. Subsequent isochronal annealings of 0.5 hour duration were done in a range $T_{ann} = 100 – 750$ °C.

Unirradiated sample of LaO$_{0.9}$F$_{0.1}$FeAs possesses metallic type conductivity in normal phase and experiences transition to superconducting state at $T_c = 30$ K. The magnetic field 13.6 T significantly reduces the superconducting transition temperature. The irradiation by neutrons suppresses superconductivity and results in complete change of temperature dependence of resistivity $\rho$. With lowering temperature $\rho$ demonstrates significant upturn. The subsequent annealing at rising temperatures in the range 100 – 300 °C does not restore the superconductivity. The annealing in the range 400 – 750 °C results in practically complete restoration of the sample's properties both in normal and superconducting phases. The evolution of resistivity $\rho$ measured at $H = 13.6$ T with neutrons irradiation and subsequent annealing at various temperatures is illustrated by the Fig. 1a.

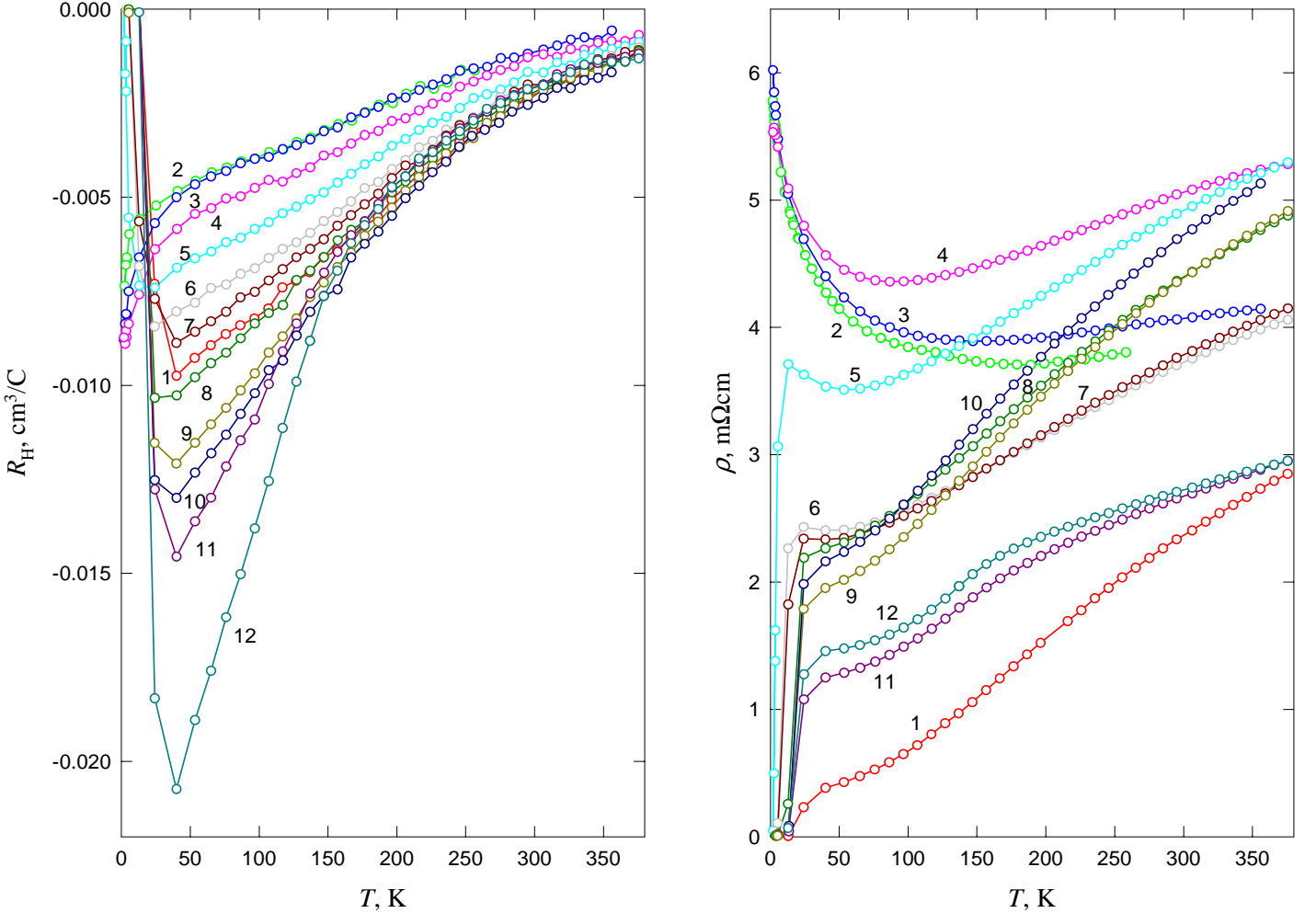

Fig. 1. Temperature dependences of Hall coefficient $R_H$ and resistivity $\rho$ in magnetic field $H = 13.6$ T of LaO$_{0.9}$F$_{0.1}$FeAs sample: initial (1), irradiated to neutron fluence $\Phi = 1.6 \cdot 10^{19}$ cm$^{-2}$ (2), and annealed at 200 °C (3), 300 °C (4), 400 °C (5), 450 °C (6), 500 °C (7), 550 °C (8), 600 °C (9), 650 °C (10), 700 °C (11) and 750 °C (12).

The temperature dependences of Hall coefficient $R_H$ in LaO$_{0.9}$F$_{0.1}$FeAs under magnetic field $H = 13.6$ T are shown in Fig. 1b. The Hall coefficient $R_H$ is negative in the normal state which means that the conduction in this system is dominated by the electron-like charge carriers. The Hall coefficient decreases with temperature down to superconducting transition temperature $T_c$ reaching value of about $10^{-2}$ cm$^3$/C. This value is in good correspondence with similar measurements in LaO$_{0.9}$F$_{0.1}$FeAs [9]. An estimation based on the single band model gives a charge carrier density of about $10^{21}$ cm$^{-3}$. The neutrons irradiation results in two times decrease of Hall coefficient which changes almost linearly down to approximately 40 K and display steeper decline below this temperature. This behavior preserves after subsequent annealing in the range 100 – 300 °C. This behavior correlates with upturn of resistivity $\rho$ in this temperature range. The annealing in the range 500 – 550 °C results in restoration of the temperature dependence of Hall coefficient of initial unirradiated sample. However, the value of superconducting transition temperature $T_c$ appears to be smaller than $T_{c0}$. Annealing in the temperature range 600 – 750 ºC leads to further increase of $R_H$, exceeding at $T = 40$ K the initial value by two times. It means that $R_H$ is not the parameter that governs $T_c$-value

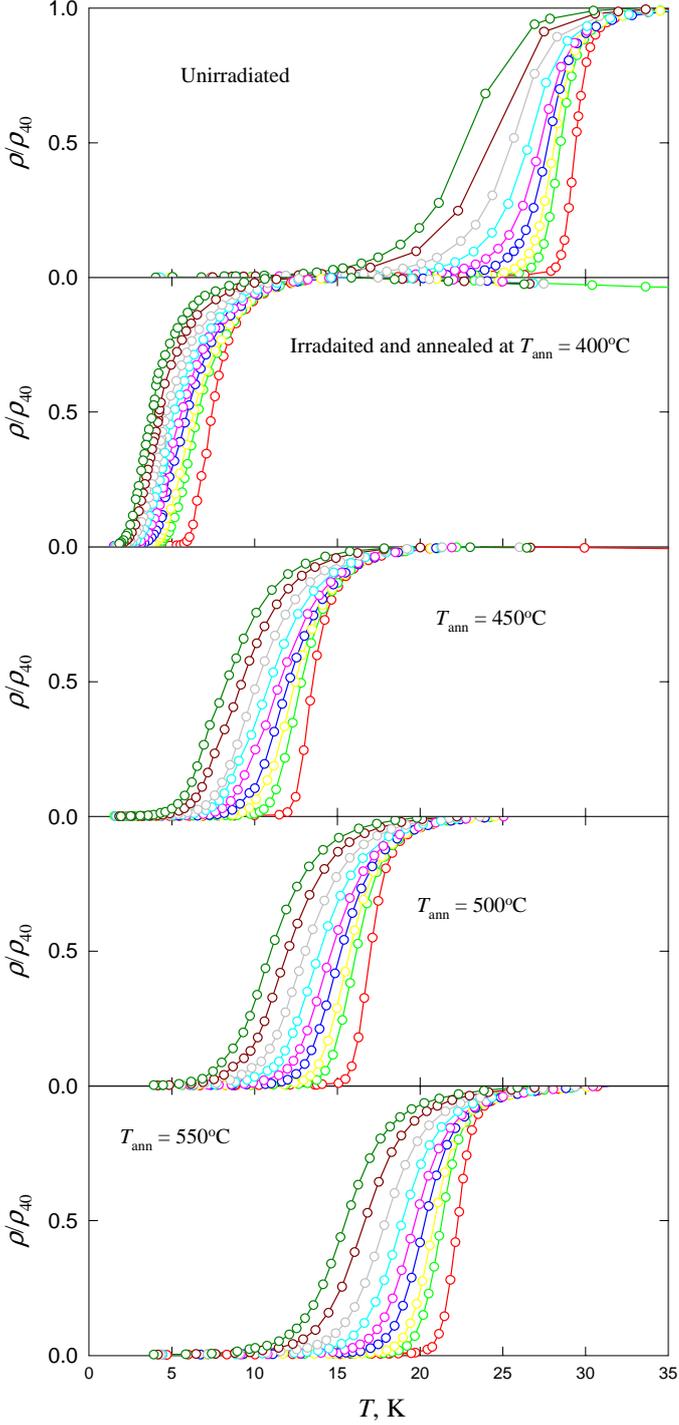

Fig. 2. Reduced (to $T = 40$ K) resistivity transition curves for initial, irradiated and annealed LaO$_{0.9}$F$_{0.1}$FeAs sample at magnetic field $H = 0$ (○), $H = 1$ T (○), $H = 2$ T (○), $H = 4$ T (○), $H = 6$ T (○), $H = 8$ T (○), $H = 10$ T (○), $H = 12$ T (○), $H = 13.6$ T (○).

Shown in Figs 2, 3 are the curves of resistivity transitions of the initial, irradiated and annealed states in magnetic fields 0, 1, 2, 4, 6, 8, 10, 12 and 13.6 T. As compared with the initial state, the irradiated and annealed (at $T_{ann} = 400$ ºC) sample features a narrower (in absolute value) superconducting transition, which is in agreement with the above suggestion on spatially uniform (here, on the coherence length scale) distribution of radiation defects. At that, transition broadening with the magnetic field increase is significantly smaller as compared with the initial sample. Further annealing leads to gradual displacement of the curves towards the initial state; however, after annealing at $T_{ann} = 750$ ºC, the value of $T_c$ is still ~2 K below $T_{c0}$.

In order to define the upper critical field slope $-dH_{c2}/dT$ from the resistivity curves of transition to a superconducting state, a criterion 0.9 or (rarely) 0.5 of the value of resistivity $\rho_n$ in the normal state is usually applied; at that, magnetoresistance noticeably affecting the accuracy of respective values definition should certainly be applied as well.

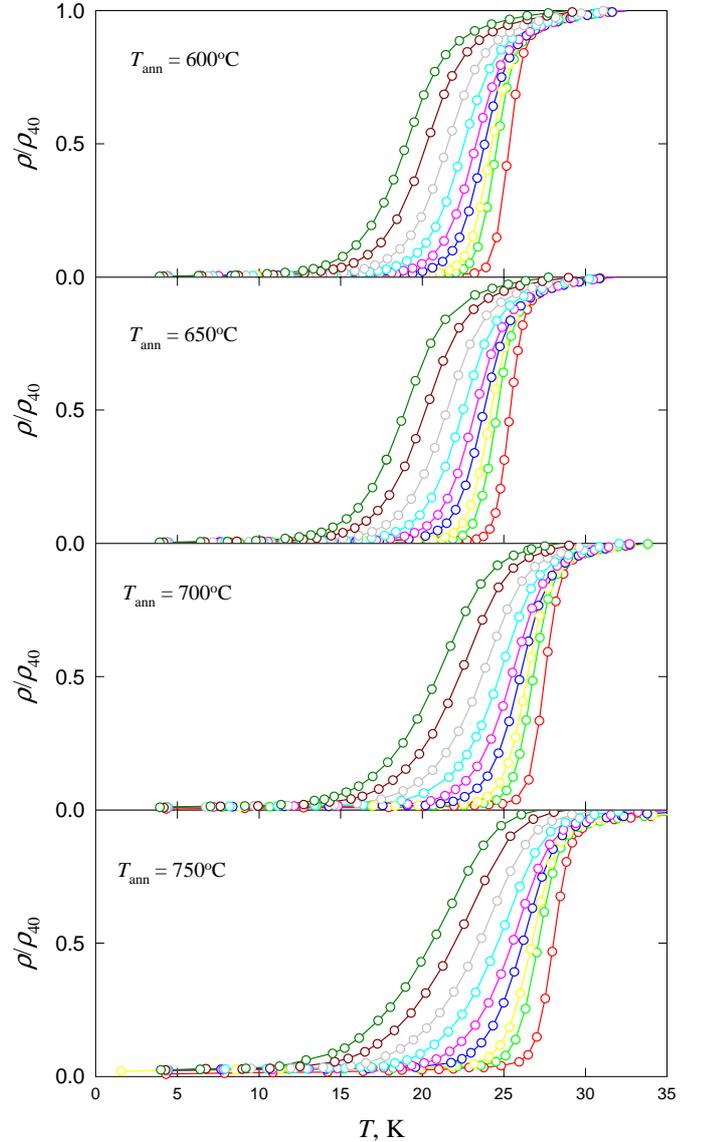

Fig. 3. Reduced resistivity transition curves for irradiated and annealed LaO$_{0.9}$F$_{0.1}$FeAs sample at magnetic field $H = 0 – 13.6$ T. Curve notations are the same as in Fig. 2.

Shown in Fig. 4 are dependences $H_{c2}(T)$ defined by these two criteria. It is worth notice that both of the above dependences are quite far from being straight lines in practically all regions of magnetic fields, particularly for the sample in the states with high values of $T_c$. The most natural explanations of such behaviour are that the initial (unirradiated) sample has material in homogeneities in composition, probably in fluorine and oxygen, as well as that resistive transition in a polycrystalline layered material is dominated by those crystallites with the *ab* planes oriented nearly perpendicular the magnetic field direction.

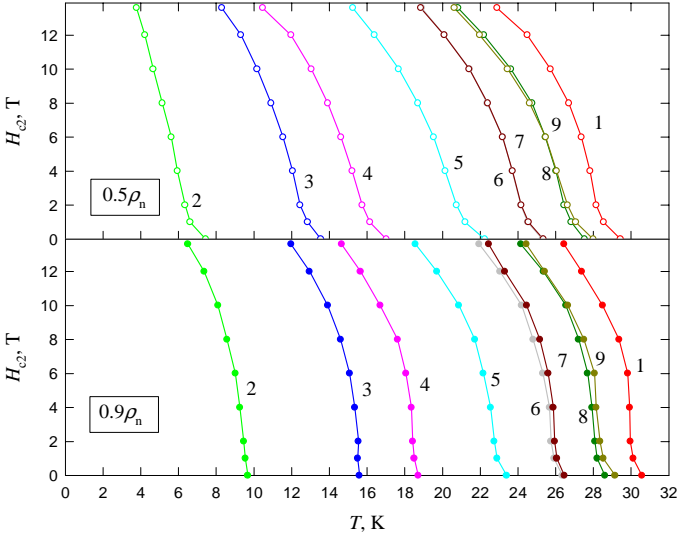

Fig. 4. Temperature dependences of upper critical field $H_{c2}$, defined at 0.9 and 0.5 of $\rho_n$ in normal state (magnetoresistance was taken into account) for initial (1), irradiated and annealed at 400 °C (2), 450 °C (3), 500 °C (4), 550 °C (5), 600 °C (6), 650 °C (7), 700 °C (8) and 750 °C (9) LaO$_{0.9}$F$_{0.1}$FeAs sample.

In the first approximation, the principal changes of $-dH_{c2}/dT$ in samples with different degrees of imperfection are nevertheless not too great, while the normal resistivity value here varied by ~15 times (from 0.3 to 4.5 mΩcm).

**Discussion**

In order to understand the causes of strong temperature dependences $R_H(T)$ and quite complex dependences $\rho(T)$ (see Fig. 1), let us consider the Hall concentration temperature dependences $n_H = 1/(R_H e)$, Fig. 5. In the temperature range of 40 K ≤ $T$ ≤ 300 K, dependences $n_H(T)$ have an exponential nature of the kind

$$n_H(T) = n_0 + n_1 \exp(-E_g/T), \qquad (1)$$

which is shown in the inset in Fig. 5, where background $n_0$ has been subtracted. Type (1) behaviour should be interpreted as excitation of charge carriers to a partially populated electron band from another band (or bands) separated by a gap of the order of $E_g$.

There are two types of systematic deviations from dependence (1) outside the range of (40 – 300) K. At low temperatures $T \leq 40$ K (for nonsuperconducting samples), there is observed a faster than almost constant decrease of $n_H(T)$ with the decrease of temperature. Even though such behaviour may be also attributed, in principle, to existence of a finer energy structure of the electron band, it will be shown in further analysis that its more correct interpretation would be the presence of a contribution from the skew Hall effect which is accounted for in Hall resistivity $\rho_{xy}$ as a sum [10]:

$$\rho_{xy} = c_H \rho M + R_H H, \qquad (2)$$

where $M$ is magnetization, $c_H$ is a constant.

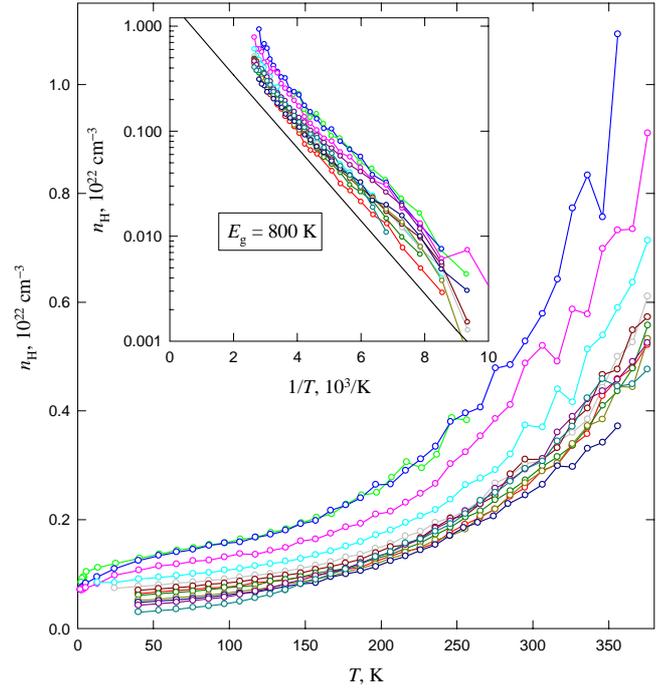

Fig. 5. Hall concentration $n_H = 1/(R_H e)$ for initial, irradiated and annealed LaO$_{0.9}$F$_{0.1}$FeAs sample as a function of temperature $T$. Insert shows $n_H$ (background is subtracted, see text) as a function of $T^{-1}$, solid line shows exponential dependence $n_H \sim \exp(-E_g/T)$ with $E_g = 800$ K. Curve notations are the same as in Fig. 1.

It may be expected that $M$ = const for metals, then at low temperatures a term with approximately logarithmic dependence will be present both in the Hall coefficient and in resistivity.

Another type of deviations from exponential dependence (1) becomes noticeable at $T > 300$ K, where $n_H$ increases much faster. It is evidently connected with the fact that at high temperatures there becomes noticeable the contribution to the Hall effect from hole-type carriers, which, for all that, have a relatively low concentration and/or low mobility. Thus Hall concentration $n_H(T)$ may, at least in the region of 40 K ≤ $T$ ≤ 300 K, be identified as the actual concentration of carriers in the electron band.

Note that such a conclusion is obviously in poor agreement with band calculations [2, 11]. Really, according to band calculations which are in good agreement with the ARPES data [12], the spectrum has two hole branches in point (0, 0) and two electron branches in points type (π, 0), the hole band top is found ~0.35 eV above the electron band bottom. At sufficiently strong electron doping, the Fermi level may lie above the hole band top, but is such case it will be difficult to expect significant increase of electrons concentration with the increase of temperature.

However, in consideration of $n_H(T)$ being an actual electron concentration, some useful parameters may be calculated. Mean free path $l^*$ for cylindrical Fermi surface model [2]

$$l^* = \{(3\pi^2)^{1/3}\hbar\{R_H\}^{2/3}\}/(\rho e^{4/3}), \quad (3)$$

where $d$ is the interplane distance, $d = 8.7$ Å. To estimate the mean free path the following circumstances are to be taken into account. The sample measured was polycrystalline with presumed porosity of about 15-20%. Due to significant anisotropy of resistivity (of about 10) in layered system, only about 1/3 of the sample participated in conductivity. According to the percolation theory [13], in the sample containing about 50% of well conducting phase the measured conductivity decreases by $k^* \approx 5$ times, so the true mean free path $l$ is connected to the value estimated from Eq. (3) via relation

$$l \cong l^* k^*. \quad (4)$$

It should be noted that, according to [14], the Hall coefficient remains practically unchanged (above the percolation threshold equal to ~ 17%).

The temperature dependences of inverse mean free path $1/l^*$ found from Eq. 2 (Fig. 6) vary quite monotonously at irradiation and annealing (compare with Fig. 1). At $T \leq 200$ K, these dependences are well described by quadratic dependences (inset in Fig. 6), which are characteristic of the prevailing electron-electron scattering

$$1/l^* = a_0 + a_2 T^2, \quad (5)$$

with a slightly faster increase at temperatures above the room temperature (similarly related to the hole band influence). At low temperatures, path $l = l^* k^*$ decreases from ~300 Å to ~12 Å, and at room temperature, the decrease is from ~15 Å to ~5 Å, which looks quite reasonable. Besides, knowing coefficient $a_2$, we may calculate effective electron mass $m^*$ [3, 15].

$$(m^*)^2 = a_2 \hbar^4 / \{2\pi V_{cell}(k_B)^2\}, \quad (6)$$

where unit cell volume $V_{cell} \approx 140$ Å$^3$. The inset in Fig. 6 shown an approximately similar slope, which yields (with account for Eq. 4) $(m^*/m_e) \approx 3$.

At low temperatures, magnetoresistance field dependences $\Delta\rho/\rho$ (Fig. 7) show a negative contribution manifestly present even in superconducting samples.

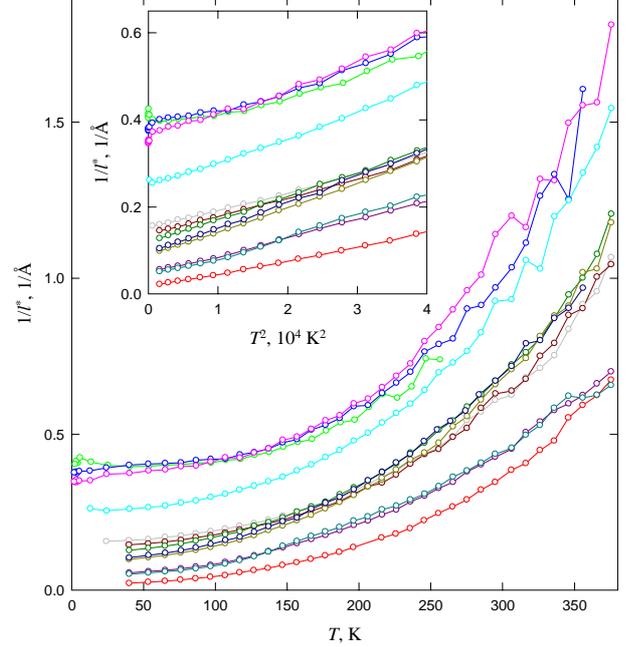

Fig. 6. Temperature dependence of inverse mean free path $1/l^*$, defined in text (Eq. 3). Inset shows $1/l^*$ as a function of $T^2$ at $T \leq 200$ K. Curve notations are the same as in Fig. 1.

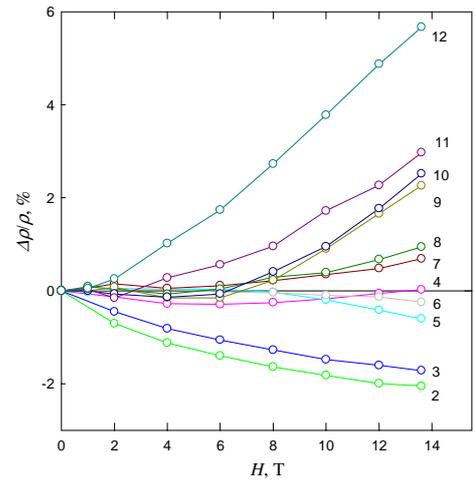

Fig. 7. Reduced magnetoresistance $\Delta\rho/\rho$ vs. $H$ at $T > T_c$ (superconducting sample) or $T = 4.2$ K (nonsuperconducting sample). Curve notations are the same as in Fig. 1.

Such behaviour of $\Delta\rho/\rho$ clearly points to significant magnetic scattering of the Kondo type, which obviously makes a logarithmic contribution to both resistivity and the Hall coefficient (skew Hall ef-fect). The value of $\Delta\rho/\rho$ as a function of a dimensionless parameter ($R_H H/\rho$) displays universal behaviour (the Kohler's rule) for the states with positive magnetoresistance.

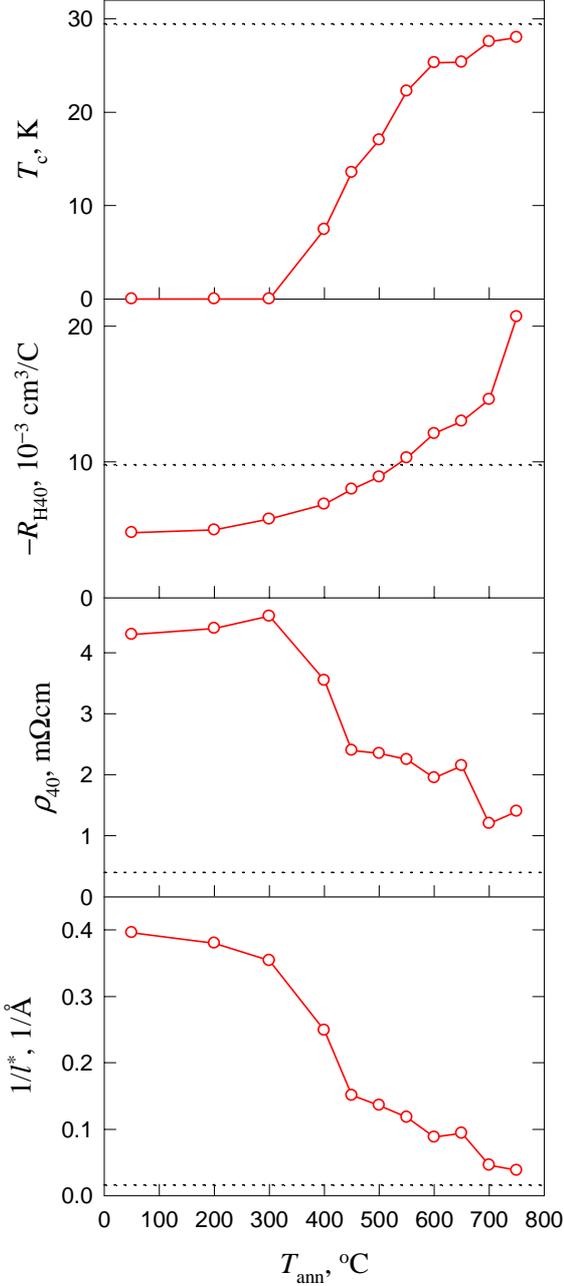

Fig. 8. Superconducting temperature $T_c$, measured at $T = 40$ K Hall coefficient $R_{H40}$ and resistivity $\rho_{40}$, extrapolated to T = 0 inverse electronic mean free path $1/l^*$ as a function of annealing temperature $T_{ann}$ for LaO$_{0.9}$F$_{0.1}$FeAs sample irradiated to neutron fluence $\Phi = 1.6 \cdot 10^{19}$ cm$^{-2}$. Dotted line shows value levels for unirradiated state.

The dependences of principal parameters of LaO$_{0.9}$F$_{0.1}$FeAs on annealing temperature $T_{ann}$ are summarized in Fig. 3. At $T_{ann} \le 300$ K, superconductivity is absent, and variation of normal state parameters is relatively small. At 300 K < $T_{ann}$ < 600 K, there takes place principal recovery of the values of $T_c$ and $1/l^*$, and at $T_{ann}$ > 600 K, these parameters vary slower. Such nonmonotonous behaviour obviously reflects the complex nature of defects emerging under irradiation and getting successively recombines in the course of annealing.

On the contrary, $R_H$ displays quite monotonous increase with the increase of $T_{ann}$ and is weakly correlated with the behaviour of $T_c$ (particularly in the region of $T_{ann} > 600$ K), which points to the important role played not by the radiation defects only, but by the effects connected with the loss of volatile components (fluorine, oxygen).

$T_c$ as a function of $1/l^*$ (and $\rho_{40}$) also displays a complicated behaviour. At low values, $1/l^* < 0.1$ ($l > 50$ Å), there is observed an approximately linear decrease of $T_c$, which further gets faster, and finally again slows down (Fig. 9). Such behaviour points to the fact that $l$ is not the only, however most significant, parameter governing the value of $T_c$.

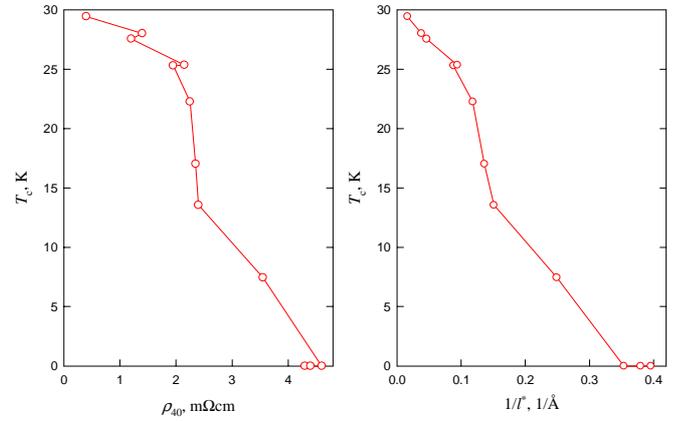

Fig. 9. $T_c$ as a function of $\rho_{40}$ and $1/l^*$ for initial, irradiated and annealed LaO$_{0.9}$F$_{0.1}$FeAs sample.

For comparison with theoretical models we use universal Abrikosov-Gor'kov (AG) equation describing suppression of superconductivity by magnetic impurities (defects) for $s$-pairing and non-magnetic ones for $d$-pairing [16]:

$$\ln(T_{c0}/T_c) = \psi(\alpha + 1/2) - \psi(1/2), \quad (7)$$

where $\alpha = \hbar/(4\pi^2 k_B T_c \tau)$, $\psi$ is digamma function, $T_{c0}$ and $T_c$ are SC temperatures of initial and disordered systems, respectively, $\tau$ is electronic relaxation time, which may be constructed from experimental values:

$$\tau = (m^* R_H)/(e\rho). \quad (8)$$

The percolation coefficient $k^*$ should be taken into account in the same way as it was done in Eq. (4).

Eq. (7) describes the decrease of $T_c$ as a function of inverse relaxation time $\tau^{-1}$, superconductivity suppressed at $\alpha > \alpha_c = 0.88$.

Fig. 10 compares in non-dimensional coordinates $T_c/T_{c0}$ vs. $\alpha = \hbar/(4\pi^2 k_B T_{c0} \tau)$ experimental dependence with AG-model. There is a good qualitatively agreement, if we consider that $\alpha$ does not contain any fitting parameters. It is difficult to expect the detail coincidence because of a number of assumptions and uncertainties, especially concerning with value of percolation coefficient $k^*$. Note that very similar behavior of superconducting properties was observed for HTSC-systems such as $YBa_2Cu_3O_7$ [17, 3], as is shown in Fig. 10.

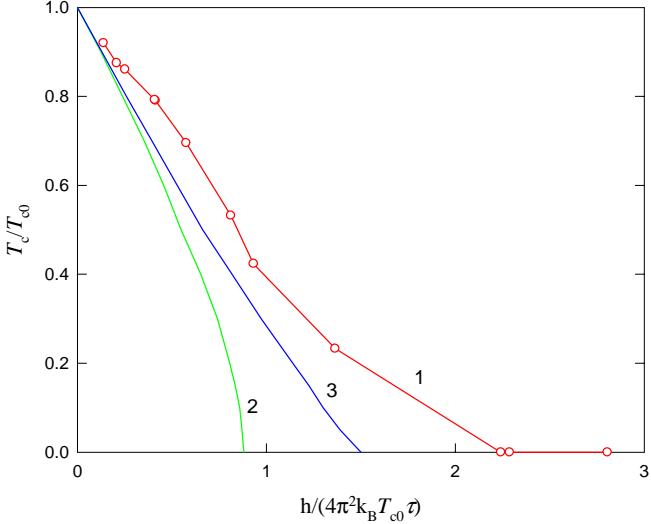

Fig. 10. $T_c/T_{c0}$ as a function of $h/(4\pi^2 k_B T_{c0} \tau)$, $T_{c0} = 32$ K. 1 – experiment, 2 – theoretical prediction, Eq. 7. Line 3 shows $T_c$-suppression of $YBa_2Cu_3O_7$ [17].

Thus the experimental data for both Cu- and Fe-based superconductors are in good agreement with the AG pair-breaking theory. It means the presence of exotic mechanism of electrons coupling in these superconductors, distinct from isotropic $s$-wave symmetry. Anisotropic $d$-pairing is realized in cuprates while isotropic pairing of anomalous $s^{\pm}$ type takes place probably in Fe-based superconductors [2]. We note the absence of any systematic study of the disordering effects in the $s^{\pm}$ type pairing model.

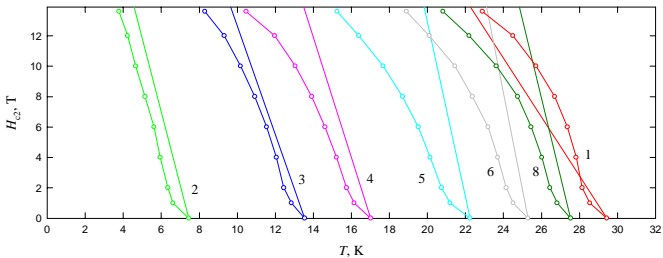

Fig. 11. Temperature dependences of upper critical field $H_{c2}$, defined at 0.5 of $\rho_n$ (redrawn from Fig. 4, circles) and calculated with Eq. 9 (straight lines).

In terms of our model values $l$ and $\tau$, we can calculate the slope of upper critical field, using standard expressions

$$-dH_{c2}/dT = \Phi_0/(0.69 \cdot 2\pi \xi^2 T_c),$$
$$1/\xi^2 = 1/\xi_0(1/\xi_0 + 1/l), \quad (9)$$
$$\xi_0 = 0.18 \, \hbar v_F/(k_B T_c).$$

The experimental and calculated values of $-dH_{c2}/dT$ are compared in Fig. 11. It seems there is no large disagreement between calculated and measured values of $H_{c2}(T)$, although accurate analysis can be done for single crystal samples only.

In conclusion, we investigated the effect of irradiation with fast neutrons and subsequent isochronal annealing on the properties of normal (resistivity $\rho(T)$, Hall coefficient $R_H(T)$) and superconducting (transition temperature $T_c$, upper critical field $H_{c2}$) states of the $LaO_{0.9}F_{0.1}FeAs$ polycrystalline sample. Fast suppression of superconductivity under irradiation point to an anomalous type of Cooper pairing in this system. The increase of radiation defects concentration (mean free path $l$ decrease) leads to an effective system doping (electrons concentration growth at their practically constant effective mass), accompanied with a monotonous decrease of $T_c$. Analysis shows that, in case of a sample with different degrees of disorder within a broad temperature range, an electron-electron scattering mechanism prevails; in the low temperature range, the Kondo type magnetic scattering makes a significant contribution.

The authors would like to thank C. Hess, R. Klingeler, B. Buechner, A. N. Vasiliev and O. S. Volkova for fruitful discussions and providing us with a high quality samples. This work was carried with partial support of the Program of Basic Research of Presidium of RAS "Condensed Matter Quantum Physics" (Project No. 4 UB RAS) and RFBR Project No. 07-02-00020-a.

Weak dependence of $R_H$ on conducting phase concentration is due to the fact that, as distinct from conductivity, on the average only the paths parallel to the measuring current contribute to the Hall voltage.